\documentclass[10pt,notitlepage]{article}
\usepackage[utf8x]{inputenc}
\usepackage[a4paper, margin=2cm]{geometry}
\usepackage{authblk}

\usepackage{amsfonts}
\usepackage{amsmath}
\usepackage{amsthm}
\usepackage{amssymb}
\usepackage{array}
\usepackage{dcolumn}
\usepackage[svgnames]{xcolor}
\usepackage[hidelinks]{hyperref}
\hypersetup{
    colorlinks=true,
    linkcolor=NavyBlue,
    filecolor=NavyBlue,
    urlcolor=NavyBlue,
    citecolor=NavyBlue,
    }

\usepackage{graphics}
\usepackage{graphicx}
\usepackage{caption}
\usepackage{subcaption}
\usepackage{adjustbox}
\usepackage{gensymb}
\usepackage{breqn}
\usepackage{braket}
\usepackage{enumitem}
\usepackage{makecell}
\usepackage{ragged2e}

\usepackage{textpos}

\def\be{\begin{equation}}
\def\ee{\end{equation}}
\def\bea{\begin{eqnarray}}
\def\eea{\end{eqnarray}}

\def\gev{\, {\rm GeV}}
\def\mev{\, {\rm MeV}}
\def\kev{\, {\rm keV}}

\def\s{\, {\rm s}}

\newcommand{\gsim}{\lower.7ex\hbox{$\;\stackrel{\textstyle>}{\sim}\;$}}
\newcommand{\lsim}{\lower.7ex\hbox{$\;\stackrel{\textstyle<}{\sim}\;$}}

\title{\large{ \bf {Opportunities for Probing $U(1)_{T3R}$ with 
Light Mediators}}}

\author[1]{\normalsize{Bhaskar Dutta\footnote{dutta@physics.tamu.edu}}}
\author[1]{\normalsize{ Sumit Ghosh\footnote{ghosh@tamu.edu}}}
\author[2]{\normalsize{Jason Kumar\footnote{jkumar@hawaii.edu}}}

\affil[1]{\small{\textit{Mitchell Institute for Fundamental Physics and Astronomy, Department of Physics and Astronomy, \protect \\ Texas A$\&$M University, College Station, Texas 77843-4242, USA }}}
\affil[2]{\small{\textit{Department of Physics and Astronomy, University of Hawaii, Honolulu, Hawaii 96822, USA}}}

\date{}

\begin{document}

\maketitle
\begin{textblock*}{3cm}(13cm,-6.5cm)
  \hbox{ {\footnotesize MI-TH-2021, UH511-1315-2020}}
\end{textblock*}

\begin{abstract}
   We consider strategies for using new datasets to probe 
   scenarios in which light right-handed SM fermions couple 
   to a new gauge group, $U(1)_{T3R}$.  This scenario 
   provides a natural explanation for the light flavor 
   sector scale, and a motivation for sub-GeV dark matter.  
   There is parameter space which is currently allowed, but 
   we find that much of it can be probed with future 
   experiments.  In particular, cosmological and 
   astrophysical observations, neutrino experiments and experiments which search for displaced visible 
   decay or invisible decay can all play a role.  Still, 
   there is a small region of parameter space which even 
   these upcoming experiments will not be able to probe.  
   This model  can explain the observed 2.4-3$\sigma$ excess of events at the
   COHERENT experiment in the parameter space allowed by current laboratory experiments, but 
   the ongoing/upcoming laboratory experiments will decisively
   probe this possibility.
\end{abstract}


\section{Introduction} \label{sec:introduction}

A well-motivated scenario for new 
physics beyond the Standard Model (BSM) is the existence 
of new gauge groups.  One well-studied gauge group, first 
considered in the context of left-right-models~\cite{Pati:1974yy, Mohapatra:1974gc, Senjanovic:1975rk}, is $U(1)_{T3R}$. In this scenario, a set of right-handed 
Standard Model (SM) fermions are charged under the new 
gauge group, while left-handed SM fermions remain uncharged.
It was recently pointed out~\cite{Dutta:2019fxn} 
that, if there is a low symmetry-breaking scale, then this scenario naturally lead to sub-GeV dark matter because the dark 
sector mass scale is tied to that of the light SM fermions.
  
There 
are a variety of 
experimental and observational
bounds on this scenario, but there is still 
open parameter space which evades all current constraints, 
and in which the dark matter candidate can achieve the 
correct relic density.  
In this work, we will consider prospects for future 
datasets to definitively probe the entirety of the 
open parameter space.

In this scenario, some 
right-handed first- and/or second-generation fermions 
are charged under $U(1)_{T3R}$, and this symmetry protects 
their masses.  If the symmetry-breaking scale is 
$\lesssim {\cal O}(10)$~GeV, then these SM 
fermions naturally obtain sub-GeV masses.  Moreover, 
this symmetry-breaking scale can naturally feed into the 
dark sector, yielding sub-GeV dark matter which interacts 
with the Standard Model through the processes mediated by 
a low-mass dark photon ($A'$) or dark Higgs ($\phi'$).

We will find that, under certain assumptions, this scenario 
can be tightly constrained by cosmological and astrophysical observations.  Collider,  fixed-target, beam-dump, 
and
coherent elastic neutrino-nucleus scattering (CE$\nu$NS) 
experiments can 
produce large fluxes of the 
dark photon and dark Higgs, which can be effectively  
probed by searching either for visible decays at distant 
detectors, or evidence for invisible decays.  Invisible 
decays of the dark photon and dark Higgs can also produce a 
flux of 
either
sterile neutrinos or of dark matter, which can 
be searched for at distant detectors 
which look for scattering.  We will see that 
models in the allowed parameter space 
can explain
the COHERENT 
excess, which is  
associated with dark matter/sterile neutrinos  emerging from the decays of 
the dark photon~\cite{Dutta:2019nbn}. Finally, this scenario can also produce non-standard interactions (NSI) of active 
neutrinos, mediated by the $A'$ or $\phi'$, which are being probed in various types of neutrino experiments.  

These new searches 
can potentially rule out or find evidence for models in 
much of the allowed parameter space.  But there are still 
regions of the parameter space which will evade bounds 
from upcoming experiments; for these models, the mediators 
decay rapidly into SM particles, leaving no signals at 
displaced detectors and signals at nearby detectors which 
are difficult to distinguish from background.

The outline of this paper is as follows.  In Sec.~\ref{sec:model}, 
we describe the model, and the interactions of the dark 
photon and dark Higgs.  In Sec.~\ref{sec:astro/cosmo}, we describe 
constraints arising from cosmological and astrophysical 
observables.  In Sec.~\ref{sec:Visible_displaced}, we describe constraints 
arising from visible decays at displaced detectors.  
We describe constraints arising from 
decays at nearby detectors in Sec.~\ref{sec:nearby detectors}.  In Sec.~\ref{sec:DMandsterile}, we 
describe constraints on dark matter or sterile neutrino 
scattering at detectors displaced from a beam source.  
In Sec.~\ref{sec:NSI}, we describe constraints on non-standard 
active neutrino interactions.  We conclude with a 
discussion of our results including a summary table in Sec.~\ref{sec:conclusion}.


\section{Model} \label{sec:model}

The details of this scenario are described in 
Ref.~\cite{Dutta:2019fxn}, but we briefly review the 
relevant properties here.
A set of right-handed Standard Model fermions are charged under $U(1)_{T3R}$, with up-type fermions having charge 
$+2$ and down-type fermions having charge $-2$.  If a 
full generation of right-handed SM fermions (including a 
right-handed neutrino) are charged under 
$U(1)_{T3R}$, then all gauge and gravitational anomalies 
cancel.  The gauge boson of $U(1)_{T3R}$ is the dark 
photon, $A'$, with gauge coupling $g_{T3R}$.

We assume that the fermions charged under $U(1)_{T3R}$ 
are first- or second-generation, with sub-GeV masses.  
Note that, to cancel anomalies, it is necessary for a 
right-handed up-type quark, down-type quark, charged 
lepton and neutrino to be charged under $U(1)_{T3R}$, but 
they need not all be in the same generation.
As shown in~\cite{Batell:2017kty}, it is technically 
natural for 
one quark (the up-type quark, 
for example) and the charged lepton to be mass 
eigenstates.  But if second-generation 
quarks are charged under $U(1)_{T3R}$, then there are 
tight constraints arising from measurements of anomalous 
Kaon decay.  We will therefore assume that the 
right-handed $u$- and $d$-quarks are charged under 
$U(1)_{T3R}$.  We will also see that there are tight 
bounds on the scenario where electrons are charged under 
$U(1)_{T3R}$, arising from cosmological 
observables~\cite{Dutta:2020jsy} and atomic parity 
violation experiments~\cite{Diener:2011jt}.  We will therefore assume that 
the right-handed muon is charged under $U(1)_{T3R}$. We also add a left-handed and right-handed fermion pair $\eta_L$ and $\eta_R$, which are 
SM singlets, but are charged under $U(1)_{T3R}$ with 
charges $\pm 1$. 
The 
$U(1)_{T3R}$ charge assignments are summarized in Table.~\ref{tab:charges}
\begin{table}[h]

\captionsetup{justification   = RaggedRight,
             labelfont = bf}
\caption{ \label{tab:charges} The 
charges of the fields which transform
under $U(1)_{T3R}$. For the fermionic fields, the shown charges are for the left-handed component of each Weyl spinor.}             
\centering
\begin{tabular}{ llllllll }
\hline\hline
field&$u_R$&$d_R$& $\mu_R$&$\nu_R$&$\eta_L$&$\eta_R$&$\phi$ \\ \hline
&&&&&&&\\
$q_{T3R}$&-2&2&2&-2&1&-1&2\\
\hline\hline
\end{tabular}
\end{table}

 With the given field content, the interaction Lagrangian can be written as,
\bea \label{intlag} \mathcal{L} = &-& \frac{\lambda_u}{\Lambda} \tilde{H} \phi^* \bar{Q}_L  u_R  -  \frac{\lambda_d}{\Lambda} H \phi \bar{Q}_L d_R - \frac{\lambda_\nu}{\Lambda} \tilde{H} \phi^*\bar{L}_L  \nu_R  - \frac{\lambda_\mu}{\Lambda} H \phi \bar{L}_L \mu_R - m_D \bar{\eta}_R  \eta_L - \frac{1}{2}\lambda_L \phi \bar{\eta}^c_L \eta_L \nonumber\\  &-& \frac{1}{2}\lambda_R \phi^* \bar{\eta}^c_R \eta_R-\mu_\phi^2 \phi^* \phi - \lambda_\phi (\phi^* \phi)^2 +H.c. , \eea where $Q_L$ and $L_L$ denote the left-handed SM quark and lepton doublet, respectively; $H$ is the SM Higgs doublet; and $\tilde{H}$ is defined as $\tilde{H}$$=$$i\tau_2H^*$.

$U(1)_{T3R}$ is broken by the condensation of a field 
$\phi$, which has charge $+2$.  This breaks $U(1)_{T3R}$ 
down to a parity, under which SM fermions are even.  
We can then express this field as 
$\phi = V + \phi' / \sqrt{2}$, where $V$ is the 
vacuum expectation value of $\phi$, and $\phi'$ is the 
dark Higgs   with mass $m_{\phi^\prime}$$=$$2\lambda_\phi^{1/2} V$.  
The mass matrix for $\eta$ 
contains both Dirac terms, $m_D$, and Majorana terms, $m_M$, the latter 
of which are necessarily proportional to $V$ as $m_M = \lambda_M V$, where we assume that $\lambda_L =\lambda_R$.  We assume 
that the Dirac terms are small compared to the Majorana 
terms, leaving us with two Majorana fermions, $\eta_1$ and 
$\eta_2$, with masses $m_1 = m_M-m_D $ and $m_2=m_M+m_D$ respectively. The mass splitting is very small, $\Delta m =2 m_D$.  The $\eta_{1,2}$ are odd under the surviving parity, 
and the lighter one, $\eta_1$ is a dark matter 
candidate.
In the low-energy effective field theory 
below the electroweak symmetry-breaking scale, SM fermions 
have  Yukawa coupling terms and mass terms of the form
\bea
{\cal L} = &-&m_u\bar{u}_L u_R  -m_d\bar{d}_L d_R  -m_{\nu D}\bar{\nu}_L \nu_R -m_\mu \bar{\mu}_L \mu_R  -\frac{1}{2} m_1 \bar{\eta}_1  {\eta}_1 -\frac{1}{2}m_2 \bar{\eta}_2  {\eta}_2-\frac{m_u }{V\sqrt{2}}\bar{u}_L u_R \phi^\prime \nonumber\\ &-&\frac{m_d  }{V\sqrt{2}}\bar{d}_L d_R \phi^\prime -\frac{m_{\nu D}  }{V\sqrt{2}}\bar{\nu}_L \nu_R \phi^\prime	-\frac{m_\mu }{V\sqrt{2}}\bar{\mu}_L \mu_R \phi^\prime  -\frac{1}{2\sqrt{2}}\frac{m_1}{V} \bar{\eta}_1  {\eta}_1 \phi^\prime-\frac{1}{2\sqrt{2}}\frac{m_2}{V} \bar{\eta}_2  {\eta}_2 \phi^\prime + H.c.  ,
\eea

The mass matrix for $\nu_{L,R}$ contains a Dirac mass 
term, $m_{\nu_D}$, which is proportional to $V$, and can contain a 
Majorana mass for $\nu_R$ which scales as $\propto 
V^2 / \Lambda$, where $\Lambda$ is some high-energy 
scale.  As such, we expect the Majorana mass to be smaller 
than $V$.  The diagonalization of the squared mass 
matrix will yield two mass eigenstates, $\nu_A$ and 
$\nu_S$.  We will assume that the active neutrino 
$\nu_A$ is mostly $\nu_L$, with only a small mixing 
of $\nu_R$.

 The interactions in the gauge sector can be explored by defining the covariant derivative as, 
\bea  D_{\mu}\mathbb{I} = {\partial}_{\mu}\mathbb{I} +i\frac{g}{2}{\tau}_a W_{\mu a}+ig^{\prime} Y B_{\mu} +i\frac{g_{T_{3R}}}{2}Q_{T_{3R}} A'_\mu. \eea where  $g$, $g^{\prime}$ and $g_{T_{3R}}$ represent the coupling constants of the  $SU(2)_L$, $U(1)_Y$ and $U(1)_{T_{3R}}$ groups, respectively. 
The respective gauge bosons are denoted by $W_{\mu }$, $ B_{\mu}$ and $A'_\mu$.  The mass of the dark photon, $A^\prime$, can be obtained from  $|D_{\mu}\phi|^2$  as $m_{A^\prime}^2=2g_{T_{3R}}^2 V^2$. The interactions involving the gauge boson $A^\prime$ are then given by,
\bea \mathcal{L}_{\text{gauge}}&=&  \frac{m_{A'}}{ 4\sqrt{2} V}A^\prime_\mu(\bar{\eta}_1\gamma^\mu\eta_2-\bar{\eta}_2\gamma^\mu\eta_1)  +  \frac{m_{A'}^2}{V\sqrt{2}} \phi' A'_\mu A'{}^{\mu} + \frac{m_{A'}^2}{4 V^2} \phi' \phi' A'_\mu {A'}^{\mu} - \frac{m_{A'}}{ 2\sqrt{2}  V}j^\mu_{A^\prime}{A^\prime}_\mu . \eea where the  interaction current for the SM fermions is defined as, $j^\mu_{A^\prime}=\sum\limits_f Q_{T_{3R}}^f \bar{f}\gamma^\mu \left( \frac{1+\gamma_5}{2} \right) f$. Note that the $\eta$ fields have only off-diagonal vector interaction with $A'$.

 All the SM fermion masses and the DM masses are proportional to the symmetry breaking scale $V$, and their masses are  $\lesssim V$. If we assume that there is no other suppression due to any other flavor physics, $V=\mathcal{O}(1)$~GeV would naturally give rise to sub-GeV masses for the fermions with $\mathcal{O}(1)$ Yukawa couplings. But this would be ruled out by the current constraints. Therefore we choose the symmetry breaking scale to be $V=10$ GeV,
leading to couplings which are moderately smaller 
than $\mathcal{O}(1)$.
The dark photon and the dark Higss masses will also be $\le \mathcal{O}(1)$~GeV.  

If $m_{\nu_S} > 2m_\mu$, then the tree-level decay process  
$\nu_S \rightarrow \mu^+ \mu^- \nu_A$ will occur rapidly.
For $m_{\nu_S} < 2m_\mu$, the sterile neutrino $\nu_S$ can decay via the process 
$\nu_S \rightarrow \nu_A \gamma \gamma$, with a rate 
\bea
\Gamma_{\nu_S} &\propto& 
\alpha_{em}^2
\frac{m_{\nu_S}^7 m_{\nu_D}^2}{m_{\phi'}^4 V^4} .
\eea

For $V=10\gev$, $m_{\phi'}\sim 100\mev$, 
$m_{\nu_S} = 10~\mev$, $m_{\nu_D} = 10^{-3}~\mev$ we find
$\tau_{\nu_S} \sim {\cal O}(10^{13})\s$.
So we may essentially assume that the light 
sterile 
neutrino is stable for the purpose of laboratory 
experiments, though it need not be 
cosmologically stable.  For 
points in parameter space at which 
the sterile neutrino is very light, it will also be a dark 
matter component. Note, the $\nu_S \rightarrow \nu_A \gamma$ decay process is also possible through a transition dipole interaction, but this process arises at two loop level and is therefore highly suppressed. 
This decay cannot proceed through 
a vector interaction, as a result of gauge invariance.


\subsection{Corrections to $g-2$}

Note that the muon anomalous magnetic moment will 
receive corrections arising from diagrams in which 
either $\phi'$ or $A'$ run in the loop. The correction to $a_\mu = (g_\mu -2)/2$  due to one-loop diagrams involving $A'$ and $\phi'$ is given by~\cite{Leveille:1977rc} 
\bea 
\label{Delta} 
\Delta a_\mu &=& \frac{m_\mu^4}{16\pi^2V^2}  \int_0^1 dx \frac{(1-x)^2(1+x)}{(1-x)^2m_\mu^2+xm_{\phi^\prime}^2} + \frac{m_\mu^2 }{32\pi^2V^2} \int_0^1 dx \frac{2x(1-x)(x-2)m_{A^\prime}^2-2x^3m_\mu^2}{x^2m_\mu^2+(1-x)m_{A^\prime}^2} .  
\nonumber\\
\eea   But it is 
important to note that $g_\mu -2$ can also receive 
corrections from high-scale physics which is disconnected 
from the dark sector.  As such, $g_\mu -2$ really constrains 
the amount of fine-tuning which is needed in order to 
match the data.  As shown in~\cite{Dutta:2019fxn}, the 
allowed parameter space for this model would require a 
fine-tuning against high-scale physics at the 1\% level 
if $V=10~\gev$.  This would be reduced to 10\% if we 
instead adopted $V=30~\gev$.


\subsection{$A'$ Interactions and Decays}  

The dark photon has a tree-level coupling to 
some right-handed SM fermions 
($u_R$, $d_R$, $\mu_R$ and  $\nu_R$), 
 with coupling strength given by,
\bea
g_{T3R} = \frac{m_{A'}}{\sqrt{2} V}.
\eea 

\begin{figure}[h]
\centering
\includegraphics[height=8cm,width=12cm]{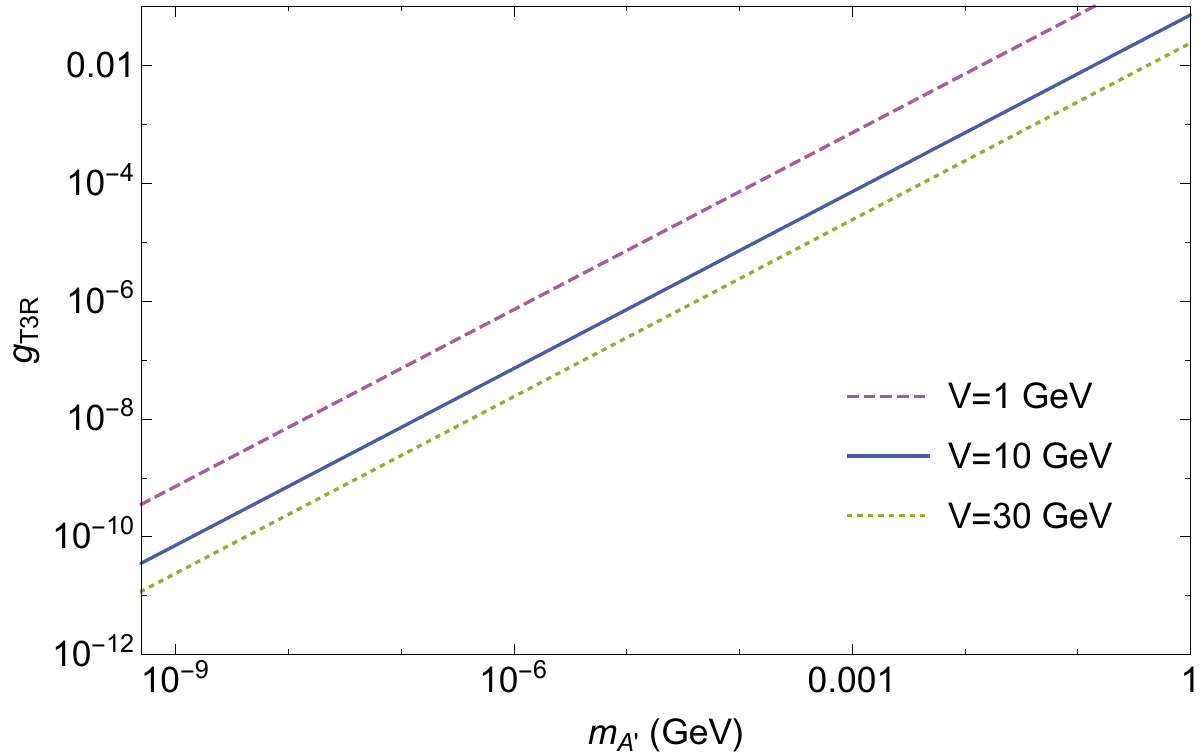}
\captionsetup{justification   = RaggedRight,
             labelfont = bf}
\caption{\label{fig:coupling}  The relation between the coupling constant $g_{T3R}$ and the gauge boson mass $m_{A'}$ for three different values of $V = 1, 10, 30$~GeV. For phenomenological study we set the value $V = 10$ GeV in rest of the paper. }
\end{figure}

The relation between $g_{T3R}$ and $m_{A'}$ is shown in Fig.~\ref{fig:coupling} for various choices of $V$.

\begin{figure}[h]
\centering
\includegraphics[height=3.5cm,width=8cm]{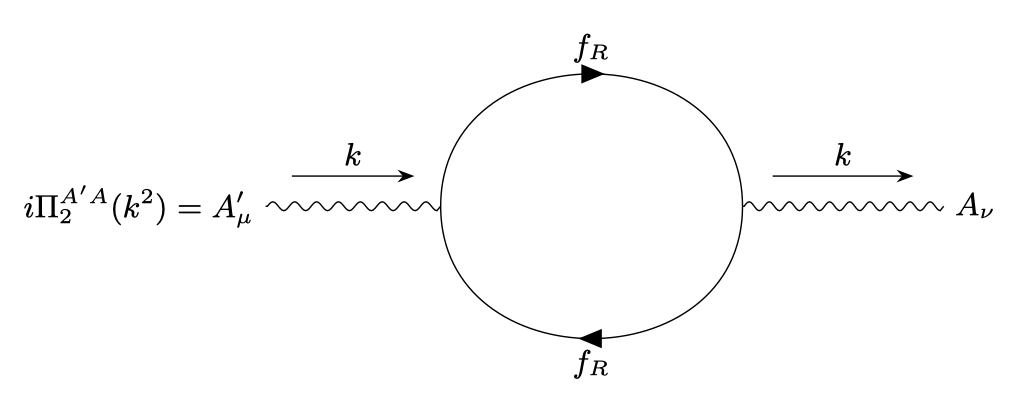}
\captionsetup{justification   = RaggedRight,
             labelfont = bf}
\caption{\label{fig:kinetic mixing}   The one loop diagrams  which give the mixing induced coupling between the SM fields to $A^\prime$. Here, $f_R= \mu_R, u_R, d_R$. }
\end{figure}

The $A'$  has a 
vector coupling to all other charged SM fermions, 
with coupling
given by $\epsilon e$, where $\epsilon$ is a 
kinetic mixing parameter.  The kinetic mixing 
parameter receives a one-loop contribution from 
the right-handed fermions charged under $U(1)_{T3R}$ 
($\sim g_{T3R} \sqrt{\alpha_{em}/ 
4\pi^3} $),
as shown in Fig.~\ref{fig:kinetic mixing}. 

But $\epsilon$ can also receive a tree-level contribution in 
the low-energy effective field theory.  One source of these contributions 
could be the integrating  out of heavy degrees of freedom which are also charged under 
$U(1)_{T3R}$.  
We will thus consider the kinetic mixing 
parameter $\epsilon$ to be a free parameter.

$A'$ also has an off-diagonal coupling to 
the vector current 
\bea
j_\eta^\mu &=& \frac{1}{2} (\bar \eta_1 \gamma^\mu \eta_2 
- \bar \eta_2 \gamma^\mu \eta_1).
\eea
Note, this coupling can only be off-diagonal, because it 
descends from a vector coupling to $\eta_{L,R}$.  This 
Dirac fermion splits into two Majorana fermions as a 
result of symmetry-breaking, but the diagonal vector 
current for a Majorana fermion vanishes identically.

Since we take the $A'$ to be lighter than $2m_\mu$ in order 
to avoid bounds from BaBar~\cite{Aubert:2009cp, Lees:2014xha}, the only potentially 
allowed two-body final states are $\eta_{1,2} \eta_{2,1}$,
$\nu \nu$, and $e^+ e^-$, of which only the last one 
is visible.  Note, the decay $A' \rightarrow \gamma \gamma$ 
is forbidden by the Landau-Yang theorem~\cite{Landau:1948kw, Yang:1950rg}.  The relevant $A'$ 
decay rates are 
\bea
\Gamma_{\eta_1 \eta_2}^{A'} &=& \frac{m_{A^\prime}^3}{96 \pi V^2} \left(1 - \frac{4m_\eta^2}{m_{A'}^2} \right)^{1/2} 
\left(1 + \frac{2m_\eta^2}{m_{A'}^2} \right) ,
\nonumber\\
\Gamma_{\nu_S \nu_S}^{A'} &=& 
\frac{m_{A'}^3}{ 12 \pi V^2}
\left(1 - \frac{4m_{\nu_S}^2}{m_{A'}^2} \right)^{3/2} 
 ,
\nonumber\\
\Gamma_{e^+ e^-}^{A'} &=& 
\frac{\epsilon^2 \alpha_{em} m_{A'}}{ 3}
\left(1 - \frac{4m_e^2}{m_{A'}^2} \right)^{1/2} 
\left(1 + \frac{2m_e^2}{m_{A'}^2} \right).
\eea
where we have assumed that $m_{\eta_2} - m_{\eta_1}$ is negligible.
Note, $A'$ has no visible two-body 
decays if $m_{A'} < 2m_e$.
If either the $\eta_1 \eta_2$ or $\nu_S \nu_S$ final states 
are kinematically allowed, then those tree-level 
decays will dominate the branching fraction.  Moreover, they will be prompt unless 
$m_{A'}$ is very small.  

If neither of those states are kinematically allowed, then 
the dominant decays will be to either $\nu_S \nu_A$, 
$\nu_A \nu_A$, or $e^+ e^-$.  The first two of these are 
suppressed by powers of the mixing angle, while the last is 
suppressed by the kinetic mixing parameter.


\subsection{$\phi'$ Interactions and Decays}

$\phi'$ couples to the SM fermions charged under 
$U(1)_{T3R}$, as well as to $\eta_1$ and 
$\eta_2$, with a coupling given by $m_f / \sqrt{2} V$.
$\phi'$ couples to $\nu_L \nu_R$ with a coupling given 
by $m_{\nu_D} / \sqrt{2} V$, where $m_{\nu_D}$ is the neutrino 
Dirac mass.

$\phi'$ can decay to $\mu^+ \mu^-$, $\eta \eta$, 
$\nu \nu$, $A' A'$ and $\gamma \gamma$.  The first and the last of these 
are visible, and the last one occurs only at one-loop.  Decays to 
$\nu_S$ or $A'$ can also produce visible energy, if those 
states in turn decay to SM particles.
Tree-level decays to hadronic states are also possible 
if $m_{\phi'} > 2m_\pi$, but 
the branching fraction to these states is 
negligible compared to $\mu^+ \mu^-$, 
because the coupling to first-generation quarks is so 
small.

The decay rates are
\bea
 \Gamma^{\phi^\prime}_{A^\prime A^\prime} &=& \frac{m_{\phi^\prime}^3}{128\pi V^2}\left( 1-\frac{4m_{A^\prime}^2}{m_{\phi^\prime}^2}\right)^{1/2} \left(1+ 12\frac{m_{A^\prime}^4}{m_{\phi^\prime}^4}-4\frac{m_{A^\prime}^2}{m_{\phi^\prime}^2} \right),
\nonumber\\
\Gamma_{\mu^+ \mu^-}^{\phi'} &=& 
\frac{m_\mu^2 m_{\phi'} }{16\pi V^2} 
\left(1 - \frac{4m_\mu^2}{m_{\phi'}^2} \right)^{3/2},
\nonumber\\
\Gamma_{\eta_i \eta_i}^{\phi'} &=& 
\frac{m_{\eta_i}^2 m_{\phi'} }{32\pi V^2} 
\left(1 - \frac{4m_{\eta_i}^2}{m_{\phi'}^2} \right)^{3/2},
\nonumber\\
\Gamma_{\nu_S \nu_A}^{\phi'} &=&  
\frac{m_{\nu_D}^2 m_{\phi'} }{16\pi V^2} 
\left(1 - \frac{m_{\nu_S}^2}{m_{\phi'}^2} \right)^{2},
\nonumber\\
\Gamma_{\gamma \gamma}^{\phi'} &=& 
\frac{\alpha_{em}^2 m_\mu^4}{8\pi^3 m_{\phi'} V^2} 
\left[1 +  \left(1 \textcolor{red}{-} \frac{4m_\mu^2}{m_{\phi'}^2} \right) 
\left(\sin^{-1} \frac{m_{\phi'}}{2m_\mu} \right)^2 \right]^2,
\nonumber\\
\eea
 where we 
have computed $\Gamma^{\phi^\prime}_{\gamma \gamma}$ only under the 
assumption $m_{\phi'} < 2m_\mu$ (otherwise, this decay 
is negligible compared to the $\mu^+ \mu^-$ channel).
Note that the decay $\phi' \rightarrow \gamma \gamma$ is always 
kinematically allowed, and (for $m_{\phi'} = 
100~\mev$) will occur at a rate 
$\sim {\cal O}(10^{12})\s^{-1}$. 

Note also that, if $m_{\phi'} > 2m_{A'}$, then 
$\phi'$ can decay promptly to $A'$.  But if this 
decay channel is dominant, then $\phi'$ production 
in a beam experiment is essentially no different 
from $A'$ production, and can be searched for using 
strategies for detecting $A'$ production.
\begin{figure*}[t]
\centering
\includegraphics[height=10.5cm,width=14.2cm]{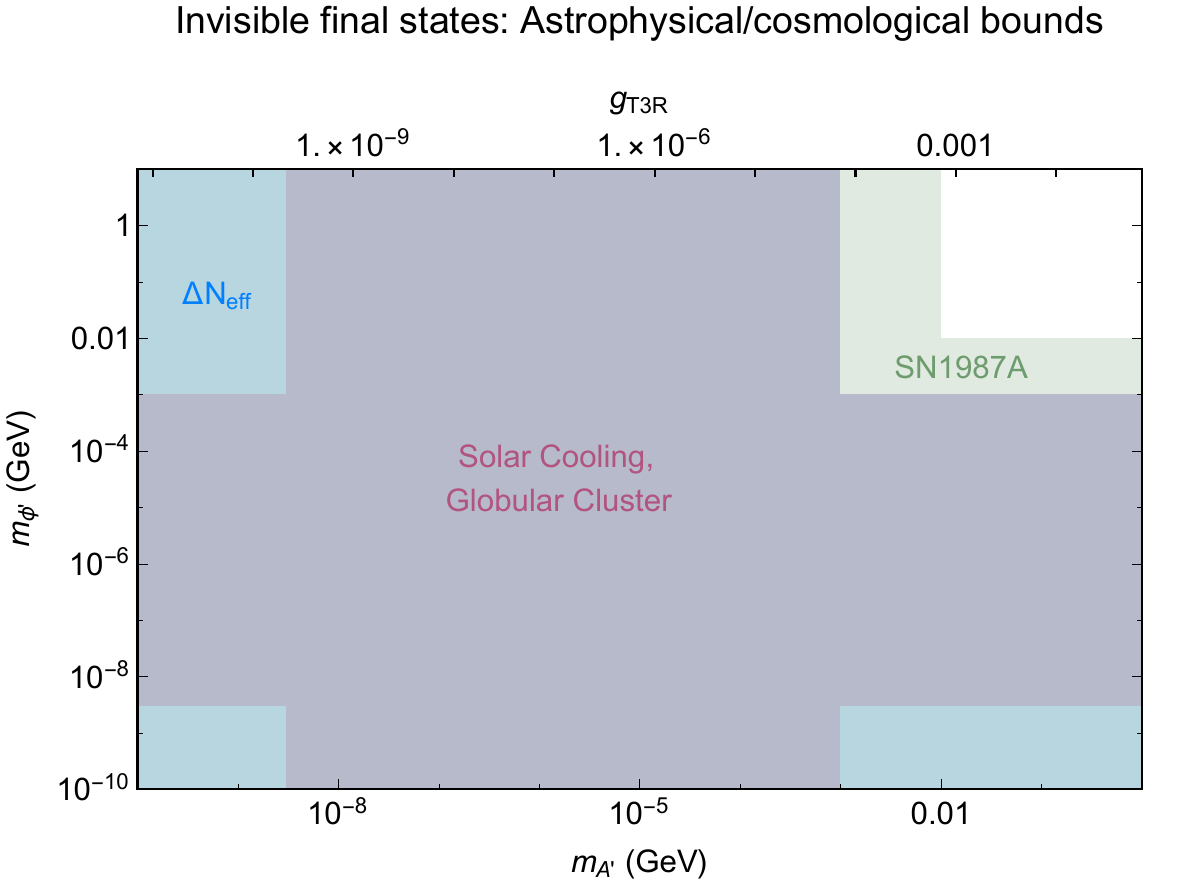}
\captionsetup{justification   = RaggedRight,
             labelfont = bf}
\caption{\label{fig:astrocosmo}  
Constraints arising from cosmological and 
astrophysical observables, assuming that the 
$A'$ and $\phi'$ decay to invisible states.  
These include constraints on 
$\Delta N_{eff}$ 
(green region) ~\cite{Dutta:2020jsy}, 
on excess cooling of stars~\cite{Harnik:2012ni, Redondo:2008aa} and globular clusters
(gray region)~\cite{Harnik:2012ni}, 
and excess cooling of supernovae (light green region)~\cite{Bollig:2020xdr,Croon:2020lrf}. The astrophysical bounds, however, are model dependent as mentioned in the text.}
\end{figure*}


\subsection{Longitudinal Polarization of $A'$}

When the gauge group is $U(1)_{T3R}$, which couples to 
chiral fermions, there is one aspect of the $A^\prime$ 
coupling to matter which is qualitatively different from 
other cases (such as $U(1)_{B-L}$, $U(1)_{L_i-L_j}$, $U(1)_{X}$~\cite{Foot:1990mn, He:1990pn, He:1991qd, Borah:2020swo, Costa:2020krs})  and which can 
have a major impact on experimental sensitivity.  
In particular, there can be an enhancement in the production 
of the $A'$ longitudinal polarization.  

The longitudinal polarization vector is $\propto 
E_{A'} / m_{A'}  \propto E_{A'} / g_{T3R} V $.  This 
yields an enhancement to the matrix element for 
processes wherein the longitudinal mode is produced 
at high-boost.  For such processes, by the Goldstone 
Equivalence theorem, the matrix element is similar to 
that for production of the Goldstone boson of $U(1)_{T3R}$ 
symmetry-breaking, with a coupling to fermions which goes as 
$m_f / \sqrt{2} V$.  For small $m_{A'}$, the coupling of a 
SM fermion to 
the longitudinal polarization is enhanced with respect to 
the transverse polarizations by a factor 
$m_f / m_{A'}$.
But this enhancement cannot be arbitrarily 
large, as it is limited by perturbative unitarity.  Since  
$m_f < V = 10\gev$, our scenario is perturbative.

Note that this enhancement only comes into play because 
the $A'$ couples to chiral SM fermions.  The vector part of 
the interaction vanishes identically for a 
longitudinally-polarized $A'$, due to the Ward identity, 
and the enhanced matrix element arises entirely from 
the axial part.  As a result, the enhancement in $A'$ 
production occurs only for chiral models such as 
$U(1)_{T3R}$, not vector-like models such as 
$B-L$, $L_i - L_j$, etc.  

As a result, the $A'$ production cross section is only 
enhanced if the $A'$ is produced at tree-level.  If 
$A'$ is produced through kinetic mixing, then the 
contribution from 
longitudinal polarization will again vanish identically 
due to the Ward Identity.  Although there is an enhancement 
of processes where $A'$ is produced through a coupling 
to $u$-/$d$-quarks, one can see from the Goldstone 
Equivalence theorem that, even for small $m_{A'}$, this 
process can be approximated by the production of a 
massless pseudoscalar with coupling $m_q /\sqrt{2} V 
\lesssim 10^{-3}-10^{-4}$.  The most dramatic effect will be 
on production of the $A'$ through a coupling to muons as  the coupling  to $u-/d-$quarks is  suppressed by close to two orders 
of magnitude, compared to the coupling to muons.  
This will be relevant for cosmological production (via 
$\mu^+ \mu^- \rightarrow \gamma A'$), production in 
supernovae (which have non-negligible muon content), 
and from future experiments involving the invisible 
decays of light $A'$ coupling directly to muons, 
such as NA64$\mu$ and LDMX-M${}^3$.


\section{Cosmological and Astrophysical Constraints} \label{sec:astro/cosmo}

There are a variety of constraints on new physics 
models which arise from cosmological and astrophysical 
observables.  

For example, if the Universe reheats to a temperature 
$\gtrsim 100~\mev$, then models with $m_{A'} < 1~\mev$ 
and $V \sim {\cal O}(10)~\gev$ 
are ruled out, because they would lead to a number of 
effective neutrinos ($N_{eff}$)~\cite{Dutta:2020jsy} which is inconsistent 
with CMB measurements~\cite{Aghanim:2018eyx}.  But this 
bound is circumvented if the Universe reheats to a 
lower temperature.  

Note that if the 
right-handed electron were charged under $U(1)_{T3R}$, 
then models with $m_{A'} < 1~\mev$ would similarly be 
ruled out by constraints on $N_{eff}$.  But this constraint 
cannot be evaded by reheating to a lower temperature; to 
avoid this constraint, one would have to reheat to a 
temperature below $1\mev$, but this is ruled out by 
BBN.

Similarly, a variety of new constraints have recently 
been shown to arise from bounds on supernovae 
cooling~\cite{Bollig:2020xdr,Croon:2020lrf}.  Essentially, 
the temperature of supernovae is large enough that a 
non-negligible population of muons is produced, and 
if they couple to new scalars or gauge bosons which 
decay invisibly, then there may be an anomalous rate 
of supernova cooling which would be ruled out by observations 
of SN1987A.  But for $m_{A',\phi'} \gtrsim 10~\mev$,  
the mediators will decay promptly, and the decay products 
will be unable to free-stream out of the 
supernova.

White dwarf (WD) cooling constraints are negligible if $m_\eta, m_{\nu_s} \ge 0.1$~MeV, in which case they will be in equilibrium with the plasmons inside the WD and can not escape. The other possible final states are $e^+e^-$ which is not allowed kinematically and $\nu_A \nu_A$, which is mixing angle suppressed~\cite{Dreiner:2013tja}. We show constraint from solar cooling~\cite{Harnik:2012ni, Redondo:2008aa} and cooling of stars in Globular clusters~\cite{Harnik:2012ni}

If Fig.~\ref{fig:astrocosmo}, we plot these cosmological and 
astrophysical bounds on the $(m_{A'}, m_{\phi'})$ 
parameter space, in the case where $A'$ and 
$\phi'$ decay invisibly.  Note that there are no
constraints plotted in the case where $m_{A'}, 
m_{\phi'} > 200 \mev$, because in this case, 
decays to $\mu^+ \mu^-$ necessarily occur at 
tree-level. 
However,  
this region of parameter 
space is already ruled out by BaBar, as 
we will see later.

 All the astrophysical constraints, however,  can be evaded  by assuming dark photon to be  chameleon-type
field with its mass  depending on the environmental matter density~\cite{Nelson:2007yq, Feldman:2006wg, Nelson:2008tn, Harnik:2012ni}.

\begin{figure}[t]
\centering
\includegraphics[height=10.5cm,width=14.2cm]{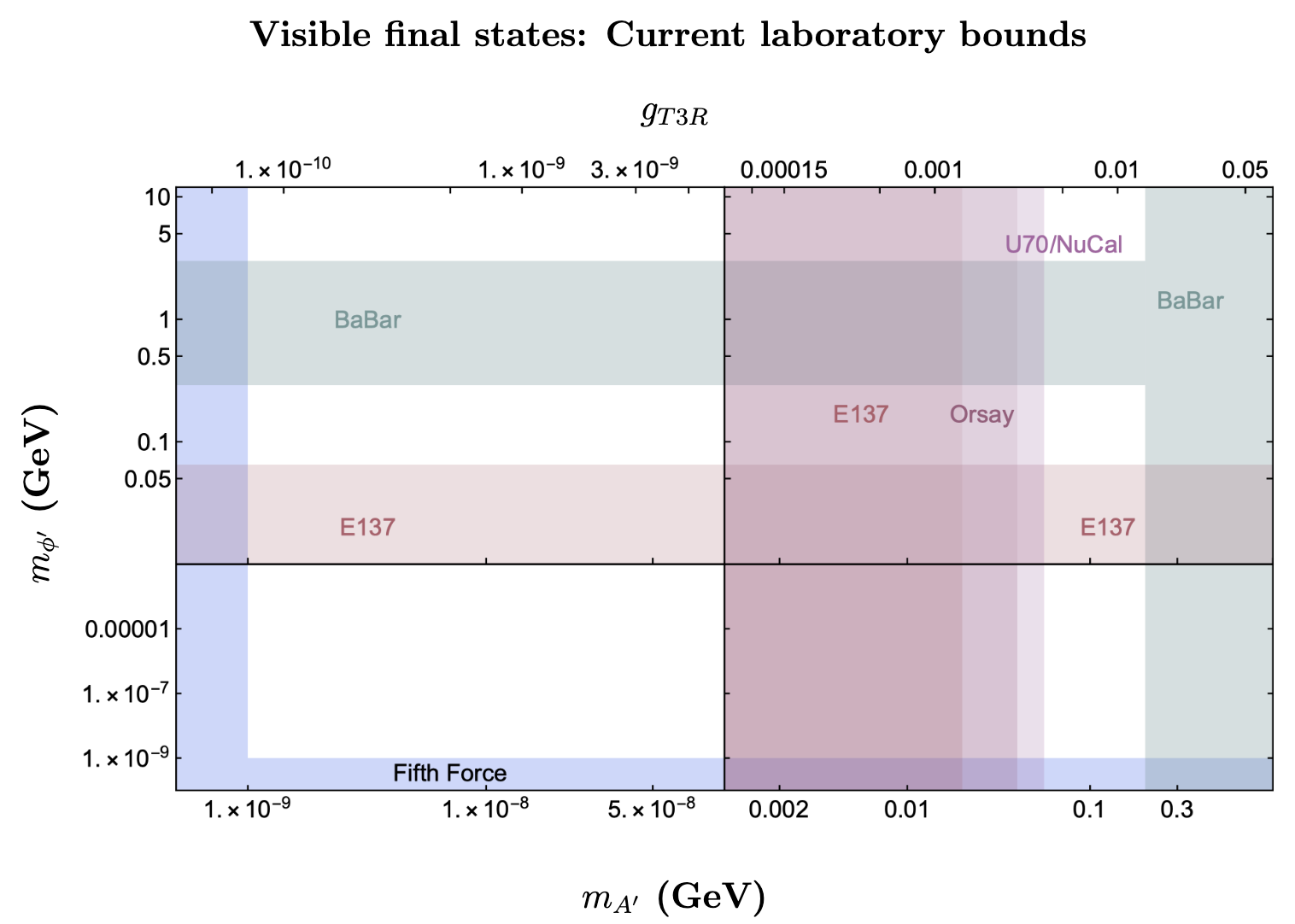}
\captionsetup{justification   = RaggedRight,
             labelfont = bf}
\caption{\label{fig:visiblecurrent} 
Regions of parameter space excluded by current 
laboratory experiments, assuming that $A'$ and 
$\phi'$ decay dominantly to SM particles.  
Included are bounds from BaBar~\cite{Aubert:2009cp, Lees:2014xha, Bauer:2018onh}, 
E137~\cite{Riordan:1987aw,Bjorken:1988as, Bjorken:2009mm}, Orsay~\cite{Davier:1989wz, Bauer:2018onh}, U70/NuCal~\cite{Gninenko:2014pea, Davier:1989wz, Bauer:2018onh} and from fifth force 
experiments~\cite{Bordag:2001qi, Harnik:2012ni}.
}
\end{figure}


\section{Visible Decays at Displaced Detectors} \label{sec:Visible_displaced}

One experimental strategy consists of producing 
light mediators at a proton 
collider, fixed-target, or beam dump experiment 
and searching for visible decays of this mediator at 
a distant detector.  Upcoming experiments of this 
type include FASER~\cite{Feng:2017uoz, Ariga:2018zuc, Ariga:2018pin, Ariga:2018uku, Ariga:2019ufm}, SHiP~\cite{Anelli:2015pba, Alekhin:2015byh}, LDMX~\cite{Mans:2017vej, Akesson:2018vlm, Moreno:2019tfm, Akesson:2019iul, Berlin:2018bsc}, and 
proposed modifications of SeaQuest~\cite{Berlin:2018pwi, Aidala:2017ofy}.

These experiments can only probe a model if the 
mediator is long-lived, and if it decays visibly. 
If the dominant decay of the $A'$ is to 
$e^+ e^-$ through kinetic mixing, then the decay length 
may be long enough for the decay to occur within the 
detector.  For the case of $\phi'$, one must determine the 
rate of $\phi'$ production at the beam, which is beyond the scope 
of this work.

For most of these experiments, 
the production of $A'$ will occur through $p$-bremsstrahlung 
and meson decay, where the $A'$ couples at tree-level to 
$u$- and $d$-quarks.  Since the $A'$ can be produced with a 
significant boost, one might wonder if the enhancement to the 
production of the longitudinal polarization will be relevant.  
But the enhancement in the coupling to the Goldstone mode, 
relative to the transverse polarizations, scales as $m_f / m_{A'}$, 
and is only large when $m_{A'}$ is small.  But visible decays are 
only possible for $m_{A'} > 1\mev$.  In any case, we will see that 
the limit of the sensitivity range for upcoming experiments will 
be $m_{A'} = {\cal O}(100)\mev$, and for such high masses, the enhancement 
to the production rate of the longitudinal polarization is minimal.  Indeed 
we will eventually see that the production enhancement for the longitudinal 
mode is most dramatic for invisible decays of light $A'$.

But although $A'$ production is a tree-level process, 
$A'$ decay occurs at one-loop through 
kinetic mixing.  As a result, the sensitivity to our model 
can be estimated by considering the estimated sensitivity 
of these experiments to models where $A'$ only couples 
to the SM via kinetic mixing, but with the number of 
events enhanced by the factor $(\pi / \alpha_{em} f)^2$, to 
account for the fact that $A'$ production is a tree-level 
process.  Here, $f$ is the factor by which the kinetic 
mixing parameter exceeds that obtained only from one-loop 
diagrams with SM fermions in the loop.

The sensitivity of displaced detector experiments is dominated 
by $A'$ produced at the largest energies, since these particles 
have the largest decay length ($\ell_{decay}$).  We will denote 
this characteristic energy as $E$, with $\ell_{decay} \propto 
(E / m_{A'}^2) \epsilon^{-2}$, assuming that the decay proceeds 
through an intermediate photon.  
We may then write $\ell_{decay}(m_{A'}) = (\epsilon_0 (m_{A'})/ \epsilon)^2 L$, where $L$ is the distance from the beam to the 
front edge of the detector, and $\epsilon_0 (m_{A'}) \propto 
m_{A'}^{-1}$ is a factor which is independent of $\epsilon$.

If $\epsilon^{kin}$ is the value of kinetic mixing parameter 
(assuming $A'$ couples to the SM only through kinetic mixing), 
then the expected number of $A'$ which decay within a detector of length 
$\Delta L$ is given by 
\bea
N_{kin} &=& C \left(\epsilon^{kin} \right)^2 
\left[e^{-(\epsilon^{kin}/\epsilon_0)^2} - 
e^{-(\epsilon^{kin}/\epsilon_0)^2 (1+\Delta L / L)}\right] ,
\eea
where $C$ is a constant which depends on the details of the 
experiment, but is independent of $\epsilon$ and $m_{A'}$.
For the dark photon of $U(1)_{T3R}$, since the $A'$ is produced 
at tree-level, the expected number of $A'$ decaying within the detector 
would instead be given by 
\bea
N_{T3R} &=& C \left(\frac{\pi}{\alpha_{em}f} \right)^2 \left(\epsilon^{T3R} \right)^2 
\left[e^{-(\epsilon^{T3R}/\epsilon_0)^2} - 
e^{-(\epsilon^{T3R}/\epsilon_0)^2 (1+\Delta L / L)}\right] ,
\eea
If we denote by $\bar N$ the number of decaying $A'$ which 
could be statistically detected above background, then the 
excluded region consists of points for which 
$N_{kin,T3R}> \bar N$.

The sensitivity of experiments of this type have a ceiling 
and floor; below the floor, the coupling is too weak for 
enough $A'$ to be produced, while above the ceiling, 
the $A'$ decays too rapidly to reach the detector.  
For a kinetic mixing parameter at the floor of 
sensitivity ($\epsilon_-$), we may assume 
$\ell_{decay} \gg L > \Delta L$, which implies that  
$(\epsilon_- / \epsilon_0)^2 (\Delta L / L) \ll 1$.

If $\epsilon_{-}^{kin} (m_{A'})$ is the floor 
of the 
sensitivity region of an experiment for the secluded 
model, then 
\bea
\bar N &=& C \left(\epsilon_-^{kin} \right)^2 
\left(\frac{\epsilon_-^{kin}}{\epsilon_0} \right)^2 
\frac{\Delta L}{L}.
\eea
The sensitivity floor of the $U(1)_{T3R}$ model 
is given by the solution to the equation
\bea
\left(\frac{\pi}{\alpha_{em}f} \right)^2 
\left(\frac{\epsilon_{-}^{T3R}}{\epsilon_{-}^{kin}} \right)^4 
&=& 1 ,
\eea
which is obtained by setting $N_{T3R} = \bar N$.

For values of kinetic mixing at the ceiling of sensitivity 
($\epsilon_+$), we may assume that $\ell_{decay} \ll \Delta L$, 
or equivalently 
$(\epsilon_+ / \epsilon_0)^2 (\Delta L / L) \gg 1$.
If $\epsilon_{+}^{kin} (m_{A'})$ is the ceiling of the 
sensitivity region of an experiment for the secluded 
model (where all dark photon interactions proceed through 
kinetic mixing), then we find 
\bea
\bar N = C \left(\epsilon_+^{kin} \right)^2 
e^{-(\epsilon_+^{kin}/\epsilon_0)^2} .
\eea
The ceiling of the $U(1)_{T3R}$ model 
is given by the solution to the equation
\bea
\left(\frac{\pi}{\alpha_{em}f} \right)^2
\left(\frac{\epsilon_+^{T3R}}{\epsilon_+^{kin}} \right)^2 
\exp \left[ \left[(\epsilon_+^{kin})^2 - ({\epsilon}_+^{T3R})^2 \right]/\epsilon_0^2 \right] &=& 1 ,
\eea
which is again obtained by setting $N_{T3R} = \bar N$.

To solve this equation, we must solve for $\epsilon_0 (m_{A'})$.  
To do this, we can use the fact that, for the secluded model, the expected
number of $A'$ decaying in the detector is $\bar N$ when the 
kinetic mixing parameter is either $\epsilon_-^{kin}$ or 
$\epsilon_+^{kin}$, yielding the relation
\bea
(\epsilon_+^{kin})^2 \exp[-(\epsilon_+^{kin})^2 / \epsilon_0^2] 
&=& \frac{(\epsilon_-^{kin})^4}{\epsilon_0^2} \frac{\Delta L}{L} .
\label{eq:VisibleDecayRelation}
\eea

Given $\epsilon_\pm^{kin} (m_{A'})$ from an experimental sensitivity or constraint 
curve, one can use the above relation to solve for 
$\epsilon_0 (m_{A'})$, and in turn determine 
$\epsilon_\pm^{T3R} (m_{A'})$.
But if visible decays are kinematically allowed 
at all, then we must have $m_{A'} > 1~\mev$; taking 
$V=10 \gev$, we find that we must have $g_{T3R} 
\gtrsim 10^{-4}$.

\begin{figure}[t]
\centering
\includegraphics[height=10.5cm,width=14.2cm]{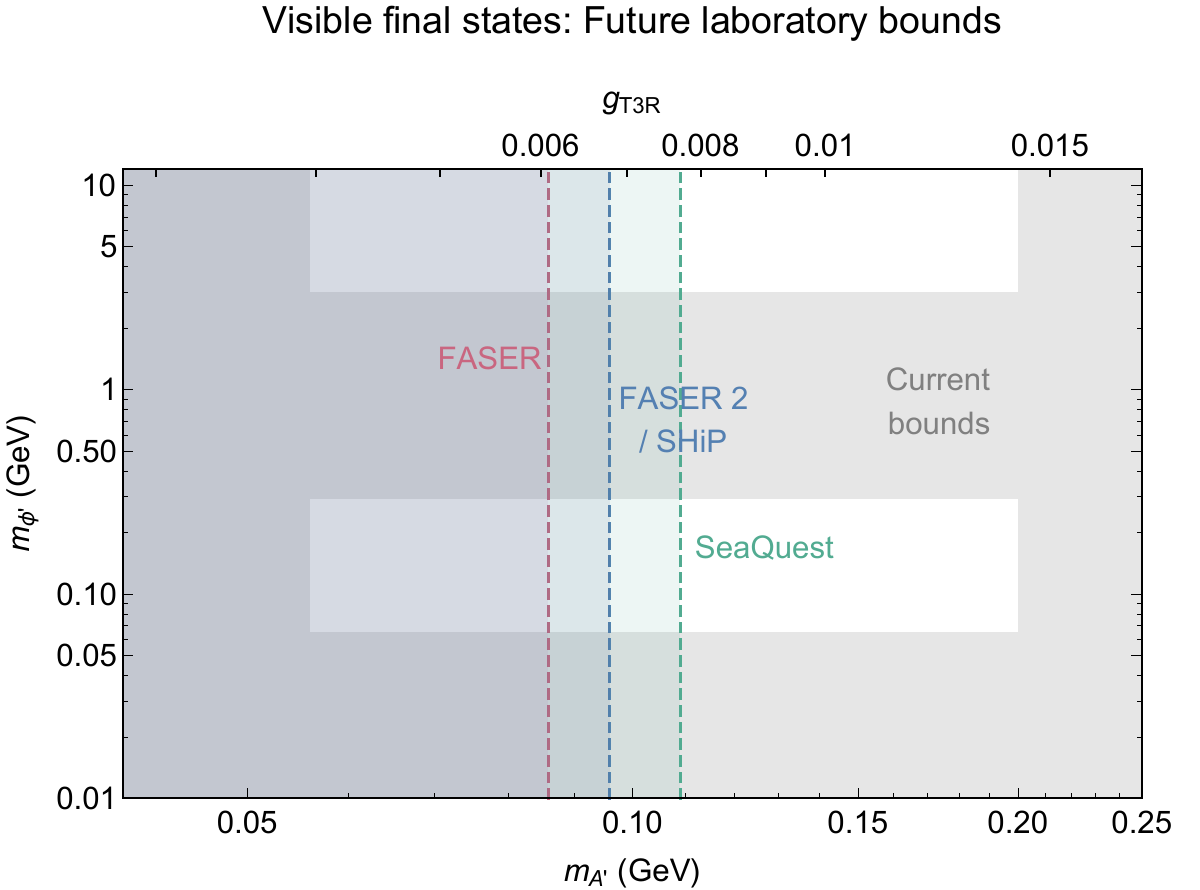}
\captionsetup{justification   = RaggedRight,
             labelfont = bf}
\caption{\label{fig:visiblefuture} The sensitivity of upcoming laboratory experiments to 
$A'$, $\phi'$ decay at displaced detectors.  Shown are the sensitivities of 
FASER~\cite{Ariga:2019ufm} (purple region), 
FASER 2/SHiP~\cite{Ariga:2019ufm} (dark green region), 
and SeaQuest~\cite{Berlin:2018pwi} 
(light green region).  Also shown are 
constraints from current laboratory experiments 
(light gray region), reproduced from 
Fig.~\ref{fig:visiblecurrent}.}
\end{figure}

If we assume that kinetic mixing is generated at one-loop 
only by SM fermions (that is, $f=1$) then we find that the 
following 
bounds (from U70/NuCal) and future
sensitivities :

\begin{itemize}
    \item{U70/NuCal: Ruled out if $1~\mev \lesssim 
    m_{A'} \lesssim 56~\mev$}
  
    \item{FASER: Probed if $1~\mev \lesssim 
    m_{A'} \lesssim 86~\mev$}
    \item{FASER-2 and SHiP (their sensitivities 
    are similar): Probed if $1~\mev \lesssim 
    m_{A'} \lesssim 96~\mev$}
    \item{SeaQuest: Probed if $1~\mev \lesssim 
    m_{A'} \lesssim 109~\mev$}
\end{itemize}
Note that, since the gauge coupling scales as 
$\propto m_{A'} / V$, increasing $V$ actually increases 
the mass reach, by increasing the decay length.

There are also 
a variety of current electron beam dump 
experiments which can constrain this 
scenario~\cite{Dutta:2019fxn}.  For these 
experiments, one-loop processes can result in the 
production of either $A'$ (kinetic-mixing) or $\phi'$ 
(Primakoff production), with subsequent one-loop decays to 
SM particles at a displaced detector.  

There are some other constraints on this scenario 
from current laboratory experiments, which were discussed in~\cite{Dutta:2019fxn}.  In particular, BaBar~\cite{Aubert:2009cp, Lees:2014xha} provides 
tight constraints on the regions of parameter 
space where $e^+ e^- \rightarrow 
\mu^+ \mu^- (A', \phi' \rightarrow \mu^+ \mu^-)$ is 
kinematically accessible.  Regions of parameter space in 
which either $\phi'$ or $A'$ are extremely light are also 
tightly constrained by fifth force experiments~\cite{Bordag:2001qi}.  Note that, 
although $g_{T3R} \propto m_{A'}$, regions of parameter 
space with very small $m_{A'}$ are still tightly constrained 
by fifth force experiments; although the transverse modes 
of the $A'$ decouple as $m_{A'} \rightarrow 0$, the 
Goldstone mode still contributes to the fifth force.

In Fig.~\ref{fig:visiblecurrent}, we plot constraints on this scenario from 
current laboratory experiments in the 
$(m_{A'}, m_{\phi'})$-plane, where we assume that $A'$, $\phi'$ predominantly 
decay to SM particles.  
In particular, we plot constraints from BaBar~\cite{Aubert:2009cp, Lees:2014xha, Bauer:2018onh}, 
E137~\cite{Riordan:1987aw,Bjorken:1988as, Bjorken:2009mm}, Orsay~\cite{Davier:1989wz, Bauer:2018onh}, U70/NuCal~\cite{Gninenko:2014pea, Davier:1989wz, Bauer:2018onh} and from fifth force 
experiments~\cite{Bordag:2001qi, Harnik:2012ni}.
 If they decay instead dominantly to 
invisible states, then these bounds are 
weakened considerably. In Fig.~\ref{fig:visiblefuture}, we plot projected bounds arising from visible decay 
at displaced detectors such as FASER, FASER-2~\cite{Feng:2017uoz, Ariga:2018zuc, Ariga:2018pin, Ariga:2018uku, Ariga:2019ufm}, SHiP~\cite{Anelli:2015pba, Alekhin:2015byh} and SeaQuest~\cite{Berlin:2018pwi, Aidala:2017ofy} in the $(m_{A'}, m_{\phi'})$-plane.


\section{Visible and Invisible Decays at Nearby detectors} \label{sec:nearby detectors}

\begin{figure}[t]
\centering
\includegraphics[height=10.5cm,width=14.2cm]{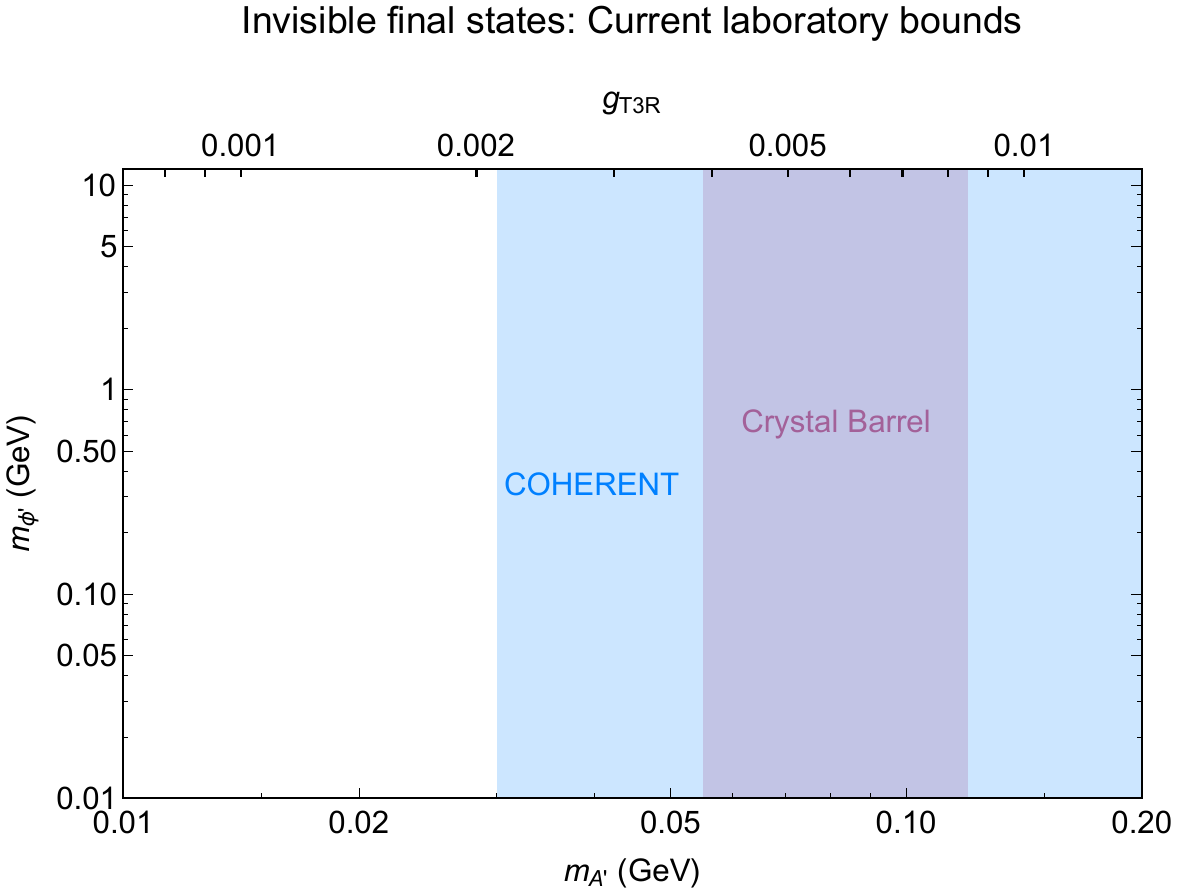}
\captionsetup{justification   = RaggedRight,
             labelfont = bf}
\caption{\label{fig:invisiblecurrent}  Regions of parameter space excluded by current 
laboratory experiments, assuming that $A'$ and 
$\phi'$ decay dominantly to invisible final states.  
Included are bounds from the COHERENT~\cite{Akimov:2017ade, Akimov:2018vzs, Akimov:2018ghi, Akimov:2019xdj, Akimov:2020pdx}(light blue) and Crystal Barrel~\cite{Amsler:1994gt, Amsler:1996hb} (light purple) experiments.}
\end{figure}

The Crystal Barrel (CB)~\cite{Amsler:1994gt, Amsler:1996hb} detector can give constraints on $m_{A^\prime}$ when $A^\prime$ predominantly decays to invisible final states. 
CB set an upper limit of the branching ratios (Br) for the process $P \rightarrow \gamma X$, where $P= \pi^0, \eta, \eta^\prime$ and $X$ is a boson which is either long-lived,  or decays invisibly. We consider  $X= A^\prime$, the dark photon.  The parameter space probed for the $\eta $ and $\eta^\prime$ decay will be ruled out by other experiments. We mainly look at $\pi^0$ decay. The bound from CB 
is~\cite{Amsler:1994gt, Amsler:1996hb}, 
    \begin{equation}
        \text{Br}(\pi^0 \rightarrow \gamma A^\prime) \le 2.8 \times 10^{-4},~~~~~~m_{A^\prime} \le 65~\text{MeV},
    \end{equation} and,
    \begin{equation}
        \text{Br}(\pi^0 \rightarrow \gamma A^\prime) \le 6.0 \times 10^{-5},~~~~~~65 ~\text{MeV}\le m_{A^\prime} \le 125~\text{MeV},
    \end{equation} 
    The branching fraction for our model is given by
    \begin{equation}
        \text{Br}(\pi^0 \rightarrow \gamma A^\prime) =  \frac{m_{A^\prime}^2}{4 \pi \alpha V^2} \left( 1- \frac{m_{A^\prime}^2}{m_{\pi^0}^2} \right)^3
    \end{equation}
    Therefore the region of paramter space, $55~ \text{MeV} < m_{A'} < 120~ \text{MeV}$ will be ruled out by CB.  

Proposed detectors such as NA64$\mu$~\cite{Chen:2017awl, Gninenko:2019qiv} and LDMX-M${}^3$~\cite{Kahn:2018cqs, Berlin:2018bsc} 
(a proposed muon beam version of 
LDMX~\cite{Akesson:2018vlm})
can probe this 
scenario in the case where either $A'$ or $\phi'$ 
has a significant decay rate to invisible states.
NA64$\mu$ proposes to collide a muon beam with a 
target, and search for interactions with missing 
energy.  In the case of a scalar mediator which 
does not decay to visible energy within the detector, 
it is estimated that NA64$\mu$ could probe muon-scalar 
couplings $\sim {\cal O}(10^{-5})$, largely independent 
of the scalar mass~\cite{Chen:2017awl}.  The coupling of $\phi'$ to muons 
is ${\cal O}(10^{-2})$, implying that this scenario can be 
probed by NA64$\mu$ for any $m_{\phi'}$, provided the 
branching fraction to invisible states satisfies 
$Br(invisible) > 10^{-6}$.  Note that, in the scenario 
in which $\phi'$ couples to muons at tree-level, LDMX-M${}^3$ 
Phase 1 will probe any  $m_{\phi'}$, provided 
$Br(invisible) > 10^{-4}$, while  
Phase 2 will have a greater 
sensitivity than 
NA64$\mu$~\cite{Berlin:2018bsc}.

Since the decay $\phi' \rightarrow \gamma \gamma$ 
is one-loop suppressed, whereas the decays 
$\phi' \rightarrow \nu_S \nu_A, \eta \eta$ 
occur at tree-level, 
these invisible decays will dominate if kinematically 
allowed.  In fact, even if the decay $\phi' \rightarrow 
\mu^+ \mu^-$ is kinematically allowed, the invisible decays 
will still have a branching fraction of at least 
${\cal O}(10^{-4})$, provided $m_{\eta, \nu_S} > 1~\mev$.  
All these scenarios can thus be probed by NA64$\mu$ and 
LDMX-M${}^3$.

Note that the sensitivity of NA64$\mu$ to $A'$ 
is roughly similar.  
Even for arbitrarily light $A'$ (with arbitrarily weak 
coupling), the longitudinal mode couples to muons 
approximately the same as a pseudoscalar with coupling 
$\sim {\cal O}(10^{-2})$.
We thus find that NA64$\mu$ and LDMX-M${}^3$
will be able to probe the 
entire parameter space, provided $m_{A', \phi'} > 
2m_\eta$ or $2\nu_S$.

Even if the $\eta \eta$ and $\nu_S \nu_S$ final states 
are not kinematically allowed, and the dominant decay of 
$A'$  is to $e^+ e^-$, NA64$\mu$ and 
LDMX-M${}^3$ will still be 
sensitive if the decay length of the $A'$ is long enough 
that a significant number of $A'$ leave the detector 
without decaying.

The sensitivity of NA64$\mu$ to a minimally flavor-violating 
(MFV) 
scalar which couples to both muons and electrons was 
considered in~\cite{Chen:2017awl}, and this case 
is essentially the same as for the $A'$.  But in that 
study it was assumed that the coupling of the 
mediator to electrons 
was suppressed relative to the coupling to muons by the 
factor $(m_e / m_\mu)$, whereas we instead assume that is 
is suppressed by the kinetic mixing factor 
$(\alpha_{em} / \pi) f$.  

As with displaced detectors, the NA64$\mu$ sensitivity 
region has a floor (below which not enough $A'$ are 
produced) and a ceiling (above which the $A'$ decays 
to visible states within the detector).  But because 
the coupling of the longitudinal polarization to muons 
is never smaller than ${\cal O}(10^{-2})$, our model 
is never below the floor for any choice of $m_{A'}.$

The ceiling of the NA64$\mu$ sensitivity to the MFV scalar 
model can be translated into a sensitivity to the 
$U(1)_{T3R}$ model by rescaling the coupling by the 
factor $(m_e / m_\mu)(\pi / \alpha_{em} f) \sim 
2.08 f^{-1}$; this 
rescaling keeps the coupling to electrons (and thus the 
decay length) fixed, while increasing the $A'$ production 
rate by an ${\cal O}(1)$ factor.  Since decreasing the 
decay length causes an exponential suppression to the 
number of events, this simple rescaling is a good approximation.
Applying this rescaling to the limits found in~\cite{Chen:2017awl}, 
we find that NA64$\mu$ is sensitive to $U(1)_{T3R}$ 
models for 
which $m_{A'} < 77~\mev$.  But if $e^+ e^-$ is the 
dominant final state, then the range 
$1\mev < m_{A'} < 56~\mev$ is already ruled out by 
U70/NuCal.  If
$m_{A'} < 1~\mev$, then no two-body 
visible decays are allowed, and both NA64$\mu$ and 
LDMX-M${}^3$ will probe this scenario.

Finally, we note that LDMX-M${}^3$ 
may have sensitivity even if dark 
matter decays to visible states within the target.  Whereas 
NA64$\mu$ relies entirely on calorimetry, LDMX relies on 
tracking, and the location of energy deposition within the 
calorimeter.  Even if prompt decays, such as $\phi' 
\rightarrow \gamma \gamma$ occur within the target, LDMX 
may be able to use information from the tracker to 
distinguish this event from Standard Model background. 
\begin{figure}[t]
\centering
\includegraphics[height=10.5cm,width=14.2cm]{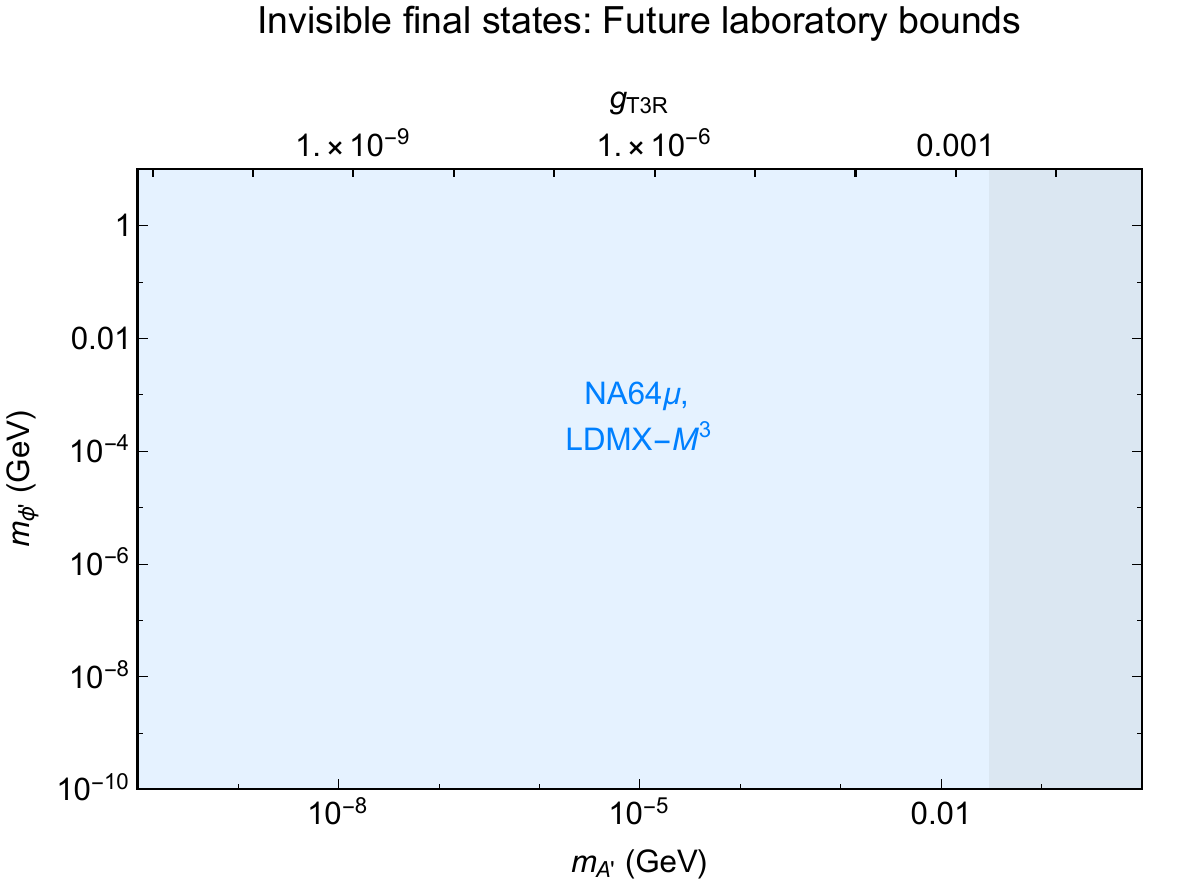}
\captionsetup{justification   = RaggedRight,
             labelfont = bf}
\caption{\label{fig:invisiblefuture} The sensitivity of upcoming laboratory experiments to the parameter space assuming that
$A'$, $\phi'$ decay dominantly to invisible final states.  Shown are the sensitivities of 
NA64$\mu$~\cite{Chen:2017awl, Gninenko:2019qiv}, 
and LDMX-$M^3$~\cite{Kahn:2018cqs, Berlin:2018bsc} 
(light blue region).  Also shown are 
constraints from current laboratory experiments 
(light gray region), reproduced from 
Fig.~\ref{fig:invisiblecurrent}.}
\end{figure}


\subsection{Belle-II: \textbf{$e^+ e^- \rightarrow 
\gamma +$ invisible }}

Since the dark photon can kinetically mix with 
the photon, Belle-II~\cite{Abe:2010gxa,Inguglia:2016acz,Bauer:2018onh} can study the process 
$e^+ e^- \rightarrow \gamma A'$.  
Its sensitivity 
is best
if the $A'$ decays invisibly~\cite{Inguglia:2016acz, Bauer:2018onh}, 
since in that case the Standard Model 
rate
is relatively small.  
For $m_{A'} \lesssim 200\mev$, Belle-II will be 
able to probe models with $g_{T3R} \gtrsim 
10^{-3}$, corresponding to $m_{A'} \gtrsim 
30 \mev$.


\section{Dark Matter and Sterile Neutrino Scattering 
at Displaced Detectors} \label{sec:DMandsterile}

There are a variety of stopped pion based experiments e.g.,  COHERENT~\cite{Akimov:2017ade, Akimov:2018vzs, Akimov:2018ghi, Akimov:2019xdj, Akimov:2020pdx}, 
CCM~\cite{CCM1, CCM2}, JSNS$^2$~\cite{Harada:2013yaa, Ajimura:2015yux, Harada:2015tcp, Ajimura:2017fld, Rott:2020duk}, etc.   which are designed to 
produce neutrinos from a proton beam hitting a target, and search for the 
neutral current scattering of these neutrinos at a 
distant detector. Among these experiments, the ongoing COHERENT and CCM experiments are CE$\nu$NS~\cite{Freedman:1973yd, Kopeliovich:1974mv} experiments. The COHERENT experiment has observed 6.7$\sigma$ (at CsI detector~\cite{Akimov:2017ade}) and 3.8$\sigma$(at LAr~\cite{Akimov:2020pdx}) evidences of CE$\nu$NS type events.  However these experiments also produce 
a tremendous number of photons from proton, electron bremsstrahlung and meson decays~\cite{Dutta:2020vop}. These photons can then produce $A'$. If the decays $A' \rightarrow \eta \eta,~ \nu_S \nu_S $ are kinematically 
allowed, then they will occur promptly and dominate 
the $A'$ branching fraction.  The scattering of the 
dark matter or sterile neutrinos against nuclei can 
then be probed. Unlike NA64$\mu$ and LDMX-M$^3$, these neutrino experiments probe the appearance of dark matter/sterile neutrinos at the detector, which makes these neutrino experiments  complimentary to the beam dump searches described before. The neutrinos from pion and muon decays produce backgrounds for such searches. However, utilizing the pulsed nature of the beam and the timing and the energy spectra of the recoiling nucleus, it has been shown 
recently 
that one can extract the dark matter signal from the neutrino background~\cite{Dutta:2019nbn}. The current result from the COHERENT experiment actually yields a tighter constraint on 
the dark matter parameter space compared to MiniBooNE~\cite{deNiverville:2011it, Aguilar-Arevalo:2017meg, Aguilar-Arevalo:2018wea, Aguilar-Arevalo:2019wki}, LSND~\cite{deNiverville:2011it, Auerbach:2001wg, Aguilar:2001ty}, NA64~\cite{Chen:2017awl, Gninenko:2019qiv} etc. In fact COHERENT CsI-data shows some excess ($\sim 2.4-3\sigma$) for $A'$ mass $\lesssim 100$ MeV where the $A'$ decays promptly to dark matter or sterile neutrino (in this model)~\cite{Dutta:2019nbn}. In Ref.~\cite{Dutta:2019nbn}, photons from $\pi^0$ decays were studied, while in Ref.~\cite{Dutta:2020vop} bremsstrahlung photons were also included to investigate dark matter emerging from 
dark photon decay, with similar results.    

The search for light dark matter/sterile neutrinos relies on kinematic features to distinguish 
the scattering of $\nu_S$ or $\eta$ from that of active 
neutrinos produced from the beam via SM processes.  
SM processes will dominantly produce active neutrinos via 
stopped pion decay, yielding neutrinos with an energy of 
$30\mev$.  But $A'$ can decay in flight to $\nu_S \nu_S$ 
or $\eta \eta$, yielding much higher energy particles.  This 
type of search was considered in detail in~\cite{Dutta:2020vop}, 
and we can apply their general results to our scenario.

Note that if sterile neutrino scattering is mediated by 
the $\phi'$, then it is exothermic, as the outgoing 
neutrino is active.  But as the sterile neutrino will already 
be boosted, the change in the event rate is an 
${\cal O}(1)$ factor.

To rescale the limits found in~\cite{Dutta:2020vop} to our 
scenario, we need only note that the dominant method for 
$A'$ production in our scenario will be from direct 
coupling to $u$- or $d$-quarks.  As such, the event rate 
is proportional to two powers of the coupling of 
$A'$ to dark matter (from the squared 
scattering matrix element) 
and four powers of the coupling of the $A'$ to first 
generation quarks (two from the squared scattering 
matrix element, and two from the squared $A'$ production 
matrix element).  
The exception is JSNS${}^2$, which looks for scattering 
against electrons; for this experiment, two powers of the 
coupling to first-generation quarks are replaced with 
the coupling to electrons.

With these rescalings, the event rate 
for the models considered in~\cite{Dutta:2020vop} can be 
directly related to the event rate in 
our scenario.  To facilitate comparison 
with~\cite{Dutta:2020vop}, we take $m_{\eta}/m_{A'}, 
m_{\nu_S}/m_{A'} = 1/3$, though deviations from this 
assumption will not affect sensitivity significantly.

If dark matter or sterile neutrino scattering is 
dominantly mediated by 
the $A'$, then from the analysis 
of~\cite{Dutta:2019nbn}, we find that the 
COHERENT excess  can be reproduced for 
$g_{T3R} \sim 0.002$.  For $V = 10\gev$, 
this corresponds to $m_{A'} \sim 30\mev$. This parameter space is not ruled out by any other available laboratory based experimental result.
Since the event rate scales as 
$g_{T3R}^6$, the dark photon mass range 
significantly above this benchmark mass is 
excluded by COHERENT data.

Note that the enhancement in coupling to the 
longitudinal mode is not relevant for 
spin-independent scattering, which is mediated 
by a purely vector interaction.

If dark matter scattering is instead mediated by the 
$\phi'$, then the event rate is suppressed by the factor 
$g_{T3R}^{-4} m_\eta^2 m_{u,d,e}^2 / 2V^4$; as long as 
$m_\eta, m_{A'} \gtrsim 30~\mev $, 
these models can also ruled out by COHERENT. 
Finally, if the 
$A'$ predominantly decays to sterile neutrinos, which 
scatter off nuclei dominantly through $\phi'$ exchange, then COHERENT rules out models for 
 which $m_{\nu_D} m_{A'} \gtrsim (30\mev)^2 $. 

Note that there are regions of parameter space 
for which these bounds can be weakened.  For 
example, changing $m_{\eta,\nu_S} / m_{A'}$ can 
effect the sensitivity by an ${\cal O}(1)$ 
factor, which could open up regions of parameter 
space at the edge of COHERENT's sensitivity.  
Similarly, if $m_{\eta,\nu_S} / m_{A'} \sim 1/2$, 
then for relatively heavy $m_{A'}$, scattering 
at the detector may be non-relativistic, leading 
to a suppression of the event rate.  
But note 
that CCM and JSNS${}^2$ will have sensitivity 
which should improve on COHERENT. This dark photon invisible decay parameter space also can be probed at DUNE~\cite{Strait:2015aku, Habig:2015rop, Abi:2020wmh, Abi:2020evt, Abi:2020oxb, Abi:2020loh}.

 Bounds from 
Crystal Barrel and COHERENT
are shown in the $(m_{A'}, m_{\phi'})$  
parameter space 
in Fig.~\ref{fig:invisiblecurrent}, assuming that the 
$A'/\phi'$ dominantly decay to invisible states. 
We also show the future sensitivities of NA64$\mu$ and LDMX-$M^3$ on the $(m_{A'}, m_{\phi'})$ parameter space in Fig.~\ref{fig:invisiblefuture}.


\section{Non-Standard Interactions for Active Neutrinos} \label{sec:NSI}

The $A'$ and $\phi'$ can mediate non-standard interactions 
of active neutrinos with nuclei.  A variety of constraints 
on such interactions have been found from oscillation 
effects, short and long baseline experiments, CE$\nu$NS experiments etc.  
Constraints can be found by using a large family of NSI, 
but considering one or two of them at a time~\cite{Giunti:2019xpr},  by reparameterizing the 
NSI down to a more phenomenological and pragmatically manageable subset based on model assumptions (for example, in Refs.~\cite{Gonzalez-Garcia:2013usa, Esteban:2018ppq, Esteban:2019lfo}), or by considering  all the NSIs from a large set  at the same time~\cite{Dutta:2020che}.  

If the momentum transfer in the scattering process is 
much smaller than the mediator mass, then the effect of the $A'$ and $\phi'$ couplings can be approximated by 
dimension-6 effective operators:
\bea
{\cal O}_{A'} &=& 
\frac{\sin^2 \theta} {2 V^2 } (\bar \nu_A \gamma^\mu P_L \nu_A) (\bar q \gamma_\mu P_R q) , 
\nonumber\\
{\cal O}_{\phi'} &=& \frac{m_q^2 \sin \theta} 
{2 V^2 m_{\phi'}^2} (\bar \nu_A  P_L \nu_A) (\bar q  P_R q),
\eea
where $\theta$ is the neutrino mixing angle.
Note that the coefficient of ${\cal O}_{A'}$ has two powers 
of the neutrino mixing angle, since $A'$ couples to 
two right-handed neutrinos, whereas the coefficient of 
${\cal O}_{\phi'}$ has only one power, as $\phi'$ couples 
to a right-handed and a left-handed neutrino.

The coefficients of these operators are bounded by current 
experiments to be of $\lesssim {\cal O}(10^{-5})~\gev^{-2}$~\cite{Dutta:2020che}, 
and future experiments could improve this bound by an 
order of magnitude.  For ${\cal O}_{A'}$, current experiments 
require $\sin^2 \theta \lesssim {\cal O}(10^{-3})$~\cite{Dutta:2020che}, 
while future experiments could probe values of 
$\sin^2 \theta$ which are an order of magnitude smaller.
For ${\cal O}_{\phi'}$, current experiments require~\cite{Dutta:2020che} 
\bea
\sin \theta \lesssim \left[ {\cal O}(10^{-3}) \right] 
\left(\frac{m_{\phi'}}{5\mev}\right)^{2} ,
\eea
with the sensitivity of upcoming experiments, e.g., DUNE~\cite{Strait:2015aku, Habig:2015rop, Abi:2020wmh, Abi:2020evt, Abi:2020oxb, Abi:2020loh}, Hyper-K~\cite{Migenda:2017oas, Yokoyama:2017mnt, Jiang:2017qhy, Bronner:2018osc, Yano:2020yfv, Kudenko:2020snj} etc. to 
$\sin \theta$ being up to an order of magnitude higher.

 The sterile neutrino, $\nu_s$, is mostly 
composed of the right-handed neutrino ($\nu_R$), 
but with a small mixing with the active neutrino.  
Because of this mixing the sterile neutrino can be produced in laboratory experiments. $\nu_s$ can be searched for 
at the HUNTER~\cite{Smith:2016vku} experiment by kinematic reconstruction of the  
electron capture decay of the radioactive atom $^{131}$Cs.  Similarly, the $\nu_s$ can potentially 
be searched for with the TRISTAN project of the  KATRIN~\cite{Mertens:2018vuu} experiment, where it can 
yield a kink-like distortion in the tritium beta decay spectrum if $m_{\nu_s} \sim \mathcal{O}(\kev)$.


\section{Conclusion}  \label{sec:conclusion}

We have considered the scenario in which right-handed 
light SM fermions are charged under a new gauge group, 
$U(1)_{T3R}$.  This scenario is of particular interest 
because it can tie the symmetry-breaking scale of 
$U(1)_{T3R}$ to that of the light SM fermions and of 
the dark sector.  This scenario thus naturally leads 
to a new set of sub-GeV particles, including the 
dark matter, a sterile neutrino, a dark photon, and 
a dark Higgs.  In this paper, we have focused on ways 
of probing this scenario with new data sets.  We have 
focused on the case where the symmetry-breaking scale 
is taken to be $10\gev$, but the results do not change 
qualitatively if that scale is increased to $\sim 
30\gev$, unless the $A'$ decays primarily to 
visible states.

We have found that the optimal probe of this scenario 
depends on the details of the model.  
Our main results are shown in 
Figures \ref{fig:astrocosmo}-\ref{fig:invisiblefuture}, 
and a summary 
of these sensitivities is presented in 
Table~\ref{table:result} .
One distinct feature of this scenario which has an 
impact on search strategies is that the dark photon 
has a chiral coupling to some SM fermions, yielding 
an enhancement to tree-level production of the 
longitudinal polarization.

If the dark photon or dark Higgs have kinematically-allowed 
decays to dark matter or to sterile neutrinos, then those 
tree-level decay processes will be prompt.  
In that case, 
excellent sensitivity arises for the ongoing 
experiments such as COHERENT, CCM and JSNS${}^2$, 
in which the $A'$ is produced at a target hit by a proton beam, 
and the invisible decay products scatter off nuclei 
in a distant detector.  
Indeed, the current $2.4-3\sigma$ excess in the event rate 
at COHERENT could be explained by a $30 \mev$ dark photon 
which is produced from photons at the target and decays to either dark 
matter or sterile neutrinos, and which also mediates 
the scattering of these invisible particles with nuclei in the target.  Constraints from COHERENT rule out 
larger masses ($m_{A'} \gtrsim 30\mev$).  
One can also find very fine-tuned regions of parameter 
space where COHERENT's sensitivity is weakened 
because the dark matter is slow-moving when it 
reaches the detector.  But CCM and JSNS${}^2$ 
will improve on the current COHERENT sensitivity. This dark photon parameter space also can be searched at DUNE.

Moreover, excellent detection prospects also 
lie with experiments 
such as NA64$\mu$ and LDMX-M${}^3$, in which a muon 
beam is collided with a target, and one searches for 
invisible decays.  These experiments can probe the 
entirety of currently available parameter space in 
which the $\phi'$ or $A'$ decay invisibly, 
including the parameter space region 
$m_{A'} \lesssim 30\mev$. The searches for dark photon at the neutrino experiments are however complimentary to the searches at NA64$\mu$ and LDMX-M${}^3$ since the former investigates the appearance of  dark matter/sterile neutrinos at the detector compared to the disappearance searches at the latter facilities. Belle-II can also 
probe models with $m_{A'} \gtrsim 30\mev$.

If the dark photon or dark Higgs decay largely to visible 
states, then the best prospects lie with beam experiments 
which search for visible decays in a distant detector, such 
as FASER, SeaQuest, and SHiP.  These models provide excellent 
detection prospects, provided the mediating particle has a 
decay length long enough to reach the detector. These experiments 
can probe dark photons in the 
${\cal O}(1-100~\mev)$ range, though there is still open 
parameter space at relatively large dark photon mass which 
these upcoming experiments cannot probe.  

Cosmological and astrophysical observables can also 
play an important role.  These constraints are especially 
interesting because, even for arbitrarily small gauge 
coupling, the longitudinal polarization of the dark 
photon has a large coupling to muons.  In particular, if 
the Universe reheats to a temperature greater than 
${\cal O}(100\mev)$, then the entire parameter space with $m_{\phi', A'} \lesssim 1\mev$ can be ruled out.  
Similarly, 
observations of SN1987A can rule 
out scenarios in either $m_{\phi', A'} \lesssim 10\mev$,
and has a dominant decay to 
dark matter or sterile 
neutrinos, though these supernovae constraints 
are subject to large systematic uncertainties, 
and can be weakened by chameleon effects, or 
other features of a more complicated dark 
sector.

We see that there is a rich and interesting 
phenomenology associated with scenarios in which 
light right-handed SM fermions are charged under 
a new gauge group, $U(1)_{T3R}$, with low-mass 
mediators.  This scenario is tightly constrained, 
yet there are still unexplored regions of 
parameter space.  

The region of parameter space 
which will not be tested with upcoming 
experiments is where the dark photon and dark 
Higgs have dominantly visible decays, but 
with a decay length which is too short to reach 
upcoming displaced detectors.  It would be 
interesting to consider new strategies for 
closing this remaining window, including 
shorter decay regions.

\begin{table}[p]
\captionsetup{justification   = RaggedRight,
             labelfont = bf}
\caption{ \label{table:result} A summary of the 
various experiments/probes considered here, their 
methods for producing and detecting the mediating particles, and the resulting sensitivities. }
\centering
\begin{adjustbox}{width=1.1\textwidth,center=\textwidth}
\begin{tabular}{ lllll }
\hline\hline

\makecell{Type of \\ experiments} & \makecell{Name of the \\ experiment} & \makecell{Production of $A^\prime/ \phi^\prime$} & \makecell{Final states} & \makecell{Results} \\
&&&& \\ \hline

\makecell{Electron \\  beam dump\\ experiments} & \makecell{E137, Orsay} & \makecell{$A^\prime$ : electron \\ bremsstrahlung\\ through kinetic \\  mixing at one-loop, \\ $\phi^\prime$ : Primakoff \\ production at \\ one-loop.} & \makecell{Both $A^\prime, \phi^\prime$ decay \\ predominantly \\ to visible SM 
\\ states $e^+e^-$. \\ $\phi^\prime$ decay is \\ rapid.} & \makecell{E137 rules out : \\ 1 MeV $\le m_{A^\prime} \le $ 20 MeV, \\ 1 MeV$\le m_{\phi^\prime} \le $ 65 MeV.\\ \\ Orsay rules out : \\ 1 MeV $\le m_{A^\prime} \le $ 40 MeV. \\} \\ 
&&&& \\

\makecell{Proton \\ beam dump \\ experiments}& \makecell{U70/NuCal, FASER \\ SHiP, SeaQuest \\ (displaced detector)} & \makecell{ $p$-bremsstrahlung \\ or meson decay \\ at tree level}& \makecell{$A^\prime \rightarrow e^+e^-$ \\ through kinetic \\ mixing. \\ $\phi^\prime \rightarrow  \gamma \gamma$ \\$\phi^\prime$ decays rapidly \\ hence cannot be probed. } & \makecell{ U70/NuCal rules out : \\ 1 MeV $\le m_{A^\prime} \le $ 93 MeV.\\  \\ FASER can probe : \\ 1 MeV $\le m_{A^\prime} \le $ 140 MeV. \\  \\ FASER 2/SHiP can probe : \\ 1 MeV $\le m_{A^\prime} \le $ 161 MeV. \\ \\  SeaQuest can probe : \\ 1 MeV $\le m_{A^\prime} \le $ 180 MeV. } \\
&&&& \\ 

\makecell{$e^+e^-$ collider \\ experiments}& \makecell{BaBar, Belle-II} &\makecell{ $e^+e^- \rightarrow \mu^+\mu^- + A^\prime/\phi^\prime $, \\ $e^+e^- \rightarrow \gamma A^\prime$}& \makecell{4$\mu$ final states, \\ $\gamma$ + invisible}& \makecell{BaBar rules out for \\ ($4\mu$ final states) : \\ 200 MeV $\le m_{A^\prime} \le $ 1.3 GeV, \\ 290 MeV $\le m_{\phi^\prime} \le $ 3 GeV. \\ \\  Belle-II can probe\\ ($\gamma$ + invisible): $m_{A^\prime} \ge 30$ MeV. } \\ 
&&&& \\ 

\makecell{$\bar{p} p$ collider \\ experiments} & \makecell{Crystal Barrel} &  \makecell{$ \bar{p} p \rightarrow \pi^0 \pi^0 \pi^0$, \\ $\pi^0 \rightarrow \gamma A'$} & \makecell{invisible states} & \makecell{ The parameter \\  space is ruled out for: \\ $55~ \text{MeV} < m_{A'} < 120~ \text{MeV}$ } \\
&&&& \\

\makecell{Fifth force \\ searches \\ experiments} & \makecell{Precision tests \\ of gravitational \\ Casimir, and \\ van der Waals forces}&\makecell{Relevant for extremely \\  light $A^\prime/\phi^\prime$. For $m_{A^\prime} \rightarrow 0$  \\ limit, the Longitudinal\\ mode  will contribute.}&\makecell{n/a}&\makecell{ The parameter \\ space is ruled out for: \\ $m_{A^\prime}/m_{\phi^\prime} \le 1$ eV. } \\
&&&& \\ 

\makecell{Astrophysical \\ probes} & \makecell{SN1987A, \\ Cooling of Sun \\ and globular clusters, \\ White dwarfs } &\makecell{ $\gamma + \mu \rightarrow A^\prime + \mu$,\\ $\mu + p \rightarrow \mu + p + A^\prime$,\\ $\mu^+\mu^- \rightarrow A^\prime$ at tree level,\\ $e^+e^- \rightarrow A^\prime$ through\\ kinetic mixing.}&\makecell{$A^\prime \rightarrow \eta \eta, \nu_s\nu_s$ (if decays to\\ $\nu \nu, e^+e^-$ then can not\\ escape),\\  $\phi^\prime \rightarrow \eta\eta,\nu \nu$ }&\makecell{SN1987A rules out : \\ $m_{A^\prime}, m_{\phi^\prime} \le 200$ MeV. \\ \\ Stellar cooling rules out:\\ $m_{A^\prime}, m_{\phi^\prime} \le 1$ MeV.\\  \\  WD constraint are negligible\\ if $m_\eta, m_{\nu_s} \ge 0.1$~MeV. \\  (All these astrophysical bounds can be\\ evaded  using chameleon effect.)} \\
&&&& \\ 

\makecell{Cosmological\\ probes} & \makecell{$\Delta N_{eff}$ value} & \makecell{ $\mu^+ \mu^- \rightarrow \gamma A^\prime$, \\ production of \\  longitudinal mode get\\ enhanced due to\\ axial vector coupling.}&\makecell{invisible states} & \makecell{If the Universe reheat at \\  a temperature $\ge 100$ MeV, \\ $m_{A^\prime}, m_{\phi^\prime} \le 1$ MeV is ruled out.\\ (Can be evaded if reheat occurs at \\a  lower temperature.)}\\ 
&&&& \\ 
\hline\hline
\end{tabular}
\end{adjustbox}
\end{table}

\begin{table}[t]
\centering
\begin{adjustbox}{width=1.1\textwidth,center=\textwidth}
\begin{tabular}{ lllll }
\hline\hline

\makecell{Type of \\ experiments} & \makecell{Name of the \\ experiment} & \makecell{Production of $A^\prime/ \phi^\prime$} & \makecell{Final states} & \makecell{Results} \\
&&&& \\ \hline

\makecell{Muon beam \\ experiments} & \makecell{NA64$\mu$, LDMX-M$^3$ \\ (nearby detectors)}& \makecell{$\mu-$bremsstrahlung}&\makecell{Can probe when \\ $A^\prime/\phi^\prime$ has a \\ significant decay rate\\ to invisible states \\ such as $\nu \nu, \eta \eta$}& \makecell{NA64$\mu$, LDMX-M$^3$ can probe \\ the entire parameter space \\ if $m_{A^\prime, \phi^\prime} > 2 m_{\eta,\nu_s}$ with \\ Br(invisible)$>10^{-4}$, \\ even if $A^\prime/\phi^\prime \rightarrow \mu^+\mu^-$ is allowed \\ still Br(invisible)$>10^{-4}$ \\ provided $m_{\eta,\nu_s} >1$ MeV. }\\ 

\makecell{Neutrino \\ 
experiments}& \makecell{COHERENT, CCM \\ JSNS$^2$} &\makecell{ $p/e$- bremsstrahlung, \\ meson decay }&\makecell{$A^\prime \rightarrow \nu_s \nu_s/\eta \eta $, \\ $\nu_s/\eta_i +N \rightarrow \nu_s/\eta_j +N$ \\generate nuclear recoil, \\ $\nu_s/\eta_i + e \rightarrow \nu_s/\eta_j + e$ \\generate electron recoil}&\makecell{Can be probed by looking at \\ nuclear/electron recoil. \\ \\ $m_{A^\prime} \sim 30$ MeV can explain the\\ 2.4-3$\sigma$ excess found by COHERENT, \\ $m_{A^\prime} \gtrsim 30$ MeV is ruled out.\\ \\  CCM and JSNS$^2$ will improve\\ the sensitivity. } \\
\hline\hline
\end{tabular}
\end{adjustbox}
\end{table}

\newpage

{\bf Acknowledgments}
We are grateful to Asher Berlin and 
Jeremy Sakstein for useful discussions. 
We acknowledge Shannon Kumar for her hospitality.
The work of BD and SG are supported in part by the DOE Grant No.~DE-SC0010813.
The work of JK is supported in part by DOE Grant No.~DE-SC0010504.

\bibliographystyle{utphys.bst}
\bibliography{dm_bound.bib}

\end{document}